\documentclass{aa}  

\usepackage{natbib}
\usepackage{graphicx}
\usepackage{txfonts}
\usepackage{hyperref}

\usepackage{color}

\begin{document} 

  \title{Machine-learning computation of distance modulus for local galaxies}

\author{
A.A. Elyiv\inst{1}
\and
O.V. Melnyk\inst{1}
\and
I.B. Vavilova\inst{1}
\and
D.V. Dobrycheva\inst{1,2}
\and
V.E. Karachentseva\inst{1}
}

\institute{Main Astronomical Observatory, National Academy of Sciences of Ukraine, 27 Akademika Zabolotnoho St., 04103, Kyiv, Ukraine\\
           \email{andrii.elyiv@gmail.com}
         \and
Bogolyubov Institute for Theoretical Physics of the NAS of Ukraine, 14-b Metrolohichna St., Kyiv, 03143, Ukraine }

  \date{Received September 30, 2019; accepted }

  \abstract
  {Quickly growing computing facilities and an increasing number of extragalactic observations encourage the application of data-driven approaches to uncover hidden relations from astronomical data. In this work we raise the problem of distance reconstruction for a large number of galaxies from available extensive observations.}
  {We propose a new data-driven approach for computing
distance moduli for local galaxies based on the machine-learning regression as an alternative to physically oriented methods.
We use key observable parameters for a large number of galaxies as input explanatory variables for training:
magnitudes in U, B, I, and K bands, corresponding colour indices, surface brightness, angular size, radial velocity, and coordinates.  } 
   {We performed detailed tests of the five machine-learning regression techniques for inference of $m-M$: linear, polynomial, k-nearest neighbours, gradient boosting, and artificial neural network regression.
As a test set we selected 91 760 galaxies at $z<0.2$ from the NASA/IPAC extragalactic database with distance moduli measured by different independent redshift methods.}
  {We find that the most effective and precise is the  neural network regression model with two hidden layers.
The obtained root-mean-square error of 0.35 mag, which corresponds to a relative error of 16\%, does not depend on the distance to galaxy and is comparable with methods based on the Tully-Fisher and Fundamental Plane relations.
The proposed model shows a 0.44 mag (20\%) error in the case of spectroscopic redshift absence and is complementary to existing photometric redshift methodologies. Our approach has great potential for obtaining distance moduli for around 250 000 galaxies at $z<0.2$ for which the above-mentioned parameters are already observed.}
  {}
{}
     {}
  

\maketitle


\section{Introduction}

Better-quality measurements of galaxy distances  than those purely dependent on redshift  are a fundamental goal in astronomy.
Such measurements are important for establishing the extragalactic distance scale, estimating the Hubble constant and cosmological models \citep{2019IJAA....9...51Z,2006FoPh...36..839H}, and studying peculiar velocities of galaxies with respect to the Hubble flow \citep{2015ApJ...805..144K,2006Ap.....49..450K,2019MNRAS.486..440D}. 
Furthermore, reconstruction of the velocity field of galaxies is crucial for mapping the Universe, and will pave the way for exploration of the large-scale structure (LSS) elements such as galaxy groups \citep{2006KFNT...22..283M,2011MNRAS.412.2498M,2012MNRAS.420.1809W, 2016RAA....16...43S, 2015MNRAS.447.2209P, 2005KFNT...21a...3V}, clusters, filaments, and voids \citep{1990ApJ...364..370B,2006MNRAS.373...45E,2012ApJ...744...43C,2015MNRAS.448..642E,2019ApJ...880...24T}, including the zone of avoidance of our galaxy \citep{2017MNRAS.471.3087S,2018RRPRA..23..244V, 2019arXiv190307461J}.

Traditionally, distances to galaxies have been measured using the distance modulus $m-M$ which is the difference between absolute and apparent stellar magnitudes.
Theoretical estimations of the absolute magnitude $M$ of a whole galaxy or some objects inside it could be performed through primary and secondary indicators.
Primary indicators are based on the standard candles, which are stars with known luminosity of which there are several types: Cepheids, RR Lyrae, Type Ia supernovae, and so on.
These methods provide distances with errors ranging from 4\% for the Local Group galaxies \citep{2012ApJ...745..156R} to 10\% for more distant galaxies.
Secondary indicators, the Tully-Fisher (TF) and Fundamental Plane (FP) empirical relationships, provide distance errors of $\sim$ 20\% and are usually applied for galaxies at $z\sim 0.1-0.2$, where individual stars are not resolved.

Despite the machine learning technique being applied in astrophysics since the 1990s \citep{1992MNRAS.259P...8S}, only in the last 10 years has computing power been sufficient to allow its widespread use \citep{2009JCAP...05..006B,2010MNRAS.408.1879N,2012JCAP...11..033N,2012cidu.conf...47V,Murrugarra-LLerenaHira:2017:GaImCl,2017arXiv171208955D,2019arXiv190407248B}.
Such a trend is also seen in rapidly increasing observational data and the development of data-driven science, where mining of datasets uncovers new knowledge.
Regression analysis occupies an important place among statistical techniques and is widely used for estimation of functional relationships between variables \citep{1990ApJ...364..104I} for spectroscopic \citep{2008A&A...477..967B} and photometric \citep{2012A&A...540A.139A} data processing. 
A machine learning genetic algorithm was used to build model-independent reconstructions of the luminosity distance and Hubble parameter as a function of redshift using supernovae type Ia data by \cite{2019arXiv191001529A,2009JCAP...05..006B}. A detailed analysis and classification of contemporary published literature on machine learning in astronomy is presented by \cite{2019arXiv191202934F}.

The redshift could be approximately reconstructed from 
the photometric redshift calculation technique (\cite{Bolzonella2000}). 
Supervised machine learning regression is commonly applied for this task \citep{2019NatAs...3..212S}.   
Sets of galaxies with multi-band photometry and known spectroscopic redshifts are used in training regression models to map
between high-dimensional photometric band space and redshifts. It is also possible to input new data into the trained model where spectroscopy is not available.
The most popular computational tools for photometric redshift inferences are the Random Forests \citep{2010ApJ...712..511C,2008ASPC..394..521C,2013MNRAS.432.1483C} and neural network regressions \citep{2012A&A...546A..13C,2015MNRAS.449.1043B} including deep machine learning techniques \citep{2018A&A...609A.111D}

Regression in a Bayesian framework was considered by \cite{2016arXiv160706059K} and was used to model the multimodal photometric redshifts. 
\cite{2019MNRAS.tmp.1905Z} published a catalogue of calibrated photometry of galaxies from the Extended Groth Strip field.
These latter authors improved photometric redshift accuracy with an algorithm based on Random Forest regression.

The principle aim of our work is to exploit available observations for galaxy datasets at redshifts $z<0.2$ in order to complement existing methods of distance measurement. For distance modulus $m-M$ (angular size distance) reconstruction we used different observational characteristics such as photometry of galaxies, their surface brightness and angular sizes, radial velocity, colour indices as analogues of morphological types, and celestial coordinates.

The influence of some parameters is not direct but we took them into account because of their possible confounding effect on $m-M$.
We also took into consideration the celestial coordinates of galaxies, which are distributed in a non-random manner in the
Universe, forming a large-scale structure web, and so we assumed that direction in the sky is important.
The probability density function (PDF) of distance to galaxy depends on the direction of observation because many galaxies are concentrated in clusters and filaments, whereas empty regions or cosmic voids occupy more than half of the volume of the Local Universe.
Previously, the galaxy coordinates were taken into account for photometric redshift computations by maximizing the
spatial cross-correlation signal between the unknown and the reference samples with redshifts \citep{2008ApJ...684...88N,2015MNRAS.447.3500R}.
\cite{2015MNRAS.454..463A} computed photometric redshifts from the product of PDFs obtained from the
colours, the cosmic web, and the local density field.

The machine learning regression approach uses a data sample for which a target value, distance modulus in our case, has already been measured with some accuracy using another direct or indirect method.
The model should be trained on a `training' sample to be able to make a regression on a new  `test' sample that has never been input into the model.
Award of training is getting minimal difference between predicted and real target value, which is an error of the model. 
Training of the model consists in  a numerical minimisation of error by changing the model parameters.
One important step is an error generalisation, where the model is evaluated on the test sample \citep{Goodfellow2016}.

We applied and compared the performances of five regression models: linear, polynomial, k-nearest-neighbour regression, 
gradient boosting, and artificial neural network (ANN).
By comparing their respective benefits and disadvantages we evaluated the $m-M$ error from the redshift-independent galaxy distance catalogue from the NASA/IPAC extragalactic database \citep{2017AJ....153...37S}.
We also considered a case where the radial velocity is not available, trying to recover $m-M$ from the available observational data.

One advantage of our approach is that we do not cut the sample by luminosity or apparent magnitude and do not impose restrictions on galaxy distribution in space 
in order to avoid losing useful information. Also, we used easily observable basic parameters, which are known for a myriad of galaxies.

In Section 2 we describe the sample of local galaxies used in this work and training sample. 
In Section 3 we discuss the main principles of machine-learning regression on the basis of linear regression.
Sections 4, 5, 6, and 7 represent the application of polynomial, k-nearest neighbour, gradient boosting, and neural network regressions, respectively. We present a discussion and our main conclusions in Section 8.


\section{Local galaxy sample}

We used the catalogue of redshift-independent distances from the NASA/IPAC extragalactic database \citep{2017AJ....153...37S}.
This is a compilation of distance measurements made using 75 different methods taken from more than 2000 refereed articles. 
The latest version 15.1.0\footnote{http://ned.ipac.caltech.edu/Library/Distances} (December 2018) contains 66388 distance measurements for 7156 galaxies. These are based on primary methods that use the standard candles such as Cepheids and Type Ia supernova, 
or standard rulers such as globular cluster radii and masers, among others. 
Also, the catalogue contains 204038 distances for 141249 galaxies based on secondary methods like TF, FP relations, and others.
The following information is available for each galaxy: identity (ID), distance modulus in mag $m-M$, one-sigma statistical error of the distance modulus, distance indicator method,
reference to a published article, reference to a published distance, and other parameters.

\begin{figure}
        \includegraphics[width=\columnwidth]{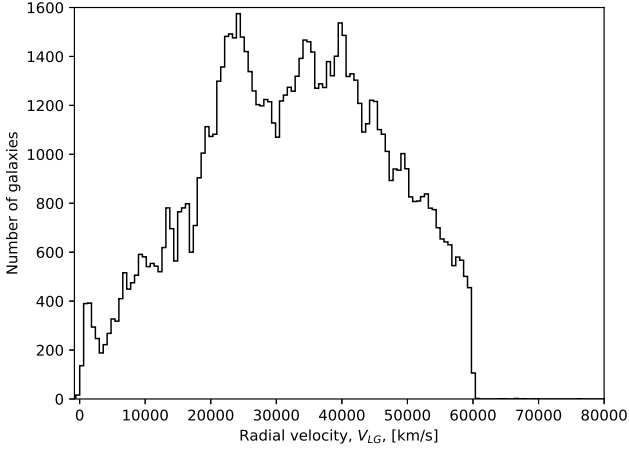}
    \caption{Radial velocity distribution of galaxies  that meet the criteria described in Section 2. The sample of galaxies is based on the catalogue of redshift-independent distances from
the NASA/IPAC extragalactic database \citep{2017AJ....153...37S}.}
    \label{fig:hyst_vlg}
\end{figure}

In order to obtain galaxy coordinates and other available observational parameters we matched them by galaxy ID with the Lyon-Meudon Extragalactic Database\footnote{http://leda.univ-lyon1.fr} (HyperLeda \cite{Mak2014}).
We considered both the northern and southern sky, except for the low galactic latitudes $|b|<15^{\circ}$ (Zone of Avoidance).
Radial velocities of galaxies were limited to $1500$ km/s $< V_{LG} < 60000$ km/s.
We did not use the nearby galaxies with $V_{LG}<1500$ km/s to avoid selection effects as the population of nearby galaxies mainly consists of dwarf galaxies (including dwarf galaxies of  low surface brightness \citep{2014Natur.513...71T,2012AstBu..67..135M, 1994BSAO...37...98K,1991MNRAS.250..802E},
which are not common among galaxy populations with $V_{LG}>1500$ km/s.
Also, the distribution of galaxies at small redshifts is very inhomogeneous due to the presence of the Virgo cluster and Tully void.
This could create a bias for the model when the galaxy coordinates are used as input explanatory variables for regression.
An upper limit at $V_{LG}=60000$ km/s was chosen since the number of galaxies with known distances drops dramatically at this velocity; see Fig.~\ref{fig:hyst_vlg}.

\begin{figure}
        \includegraphics[width=\columnwidth]{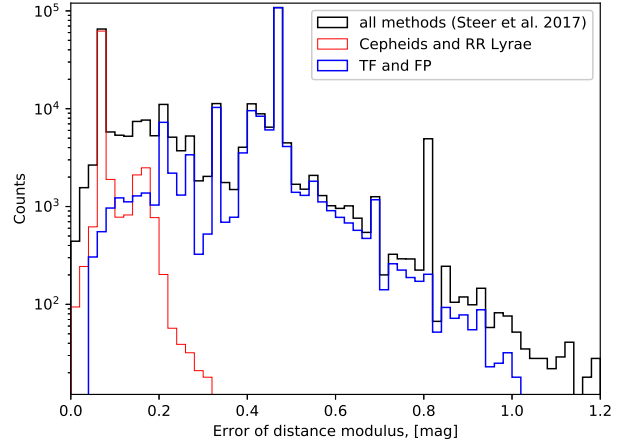}
    \caption{Distribution of distance modulus errors for all galaxies from Steer et al. 2017 (black line). 
    As an example, the distributions are shown for typical methods: Cepheids and RR Lyrae (red thin line), TF and FP relations methods (blue thick line).}
    \label{fig:hyst_err}
\end{figure}

We reduced all of the distance moduli to the common Hubble constant $H_{0}=70~$km~s$^{-1}~$Mpc$^{-1}$, 
since this is the default value used by the Supernova Cosmology Project and the Supernova Legacy Survey \citep{2017AJ....153...37S}.
Distance modulus error depends on the method used  and varies in a wide range from 0.06 mag for Cepheids and RR Lyrae Stars to 0.42 mag for the FP and TF methods (Fig.~\ref{fig:hyst_err}). We used measurements made with primary\footnote{Maser, SNII optical, the old globular cluster luminosity function method (GCLF), planetary nebulae luminosity function (PNLF),  Surface Brightness Fluctuations (SBF) method,
Miras, SNIa SDSS, TypeII Cepheids, SNIa, HII region diameter, Horizontal Branch, colour-magnitude diagram (CMD), Eclipsing Binary, Tip of the Red Giant Branch (TRGB), RR Lyrae, Cepheids, Red Clump} and 
secondary\footnote{FP, Brightest Stars, TF, Tertiary, D-Sigma, Brightest Cluster Galaxy (BCG), colour-magnitude diagram (CMD), Faber-Jackson, Active galactic nucleus (AGN) time lag, Sosies, Magnitude}
methods with a mean error of less than 0.50 mag to train our models.
However, all the individual measurements with error above 0.50 mag were removed from the sample. 
Some galaxies have several distance measurements taken by different authors using different methods. We aggregated such distances for each galaxy and calculated the weighted mean $m-M$ with the weight inversely proportional to the square of the error.

Finally, we obtained an initial sample of 91760 galaxies, $S_{0.50}$, with the following information: 
Supergalactic coordinates (SGB, SGL); radial velocity with respect to the Local Group, $V_{LG}$; the decimal logarithm of the projected major axis length
of the galaxy at the isophotal level of 25 mag/arcsec$^{2}$ 
in the B-band $logd25$; mean surface brightness within 25 mag isophote in the B-band, $bri25$; the apparent total U, B, I, and K magnitudes; and U-I and B-K colours, which represent a morphological type of galaxy. We did not use other observational parameters like 21cm line flux and velocity rotation since these are only available for < 3\% galaxies in our sample.

We created the second sample of galaxies limiting with $m-M$ error of $<$ 0.25 mag and $1500$ km/s $< V_{LG} < 30000$ km/s. 
This sample contains 9360 galaxies with distances calculated mostly using accurate Cepheids and RR Lyrae methods.
We refer to this sample as $S_{0.25}$. 

For the correct application of the machine-learning algorithms we reduced $V_{LG}$ to magnitude units using the following conversions.

\begin{equation}
m-M=5*log\left(\frac{d_{A}}{10pc}\right),
        \label{eq:mM}
\end{equation}

where angular size distance for the ${\Lambda}$CDM cosmological model:
\begin{equation}
d_{A}=cH_{0}^{-1}(1+z)^{-1}\int_{0}^{z'}{dz'/\sqrt{\Omega_{m}(1+z')^{3}+\Omega_{\Lambda}}},
        \label{eq:da}
\end{equation}
with parameters $H_{0}=70~$km~s$^{-1}~$Mpc$^{-1}$, $\Omega_{m}=0.3089$, $\Omega_{\Lambda}=0.6911$.
Finally, the redshift could be expressed as a radial velocity $V_{LG}$ as:
\begin{equation}
1+z=\sqrt{\frac{1+V_{LG}/c}{1-V_{LG}/c}}.
\label{eq:z}
\end{equation}
We do not consider the K-correction here because it is significantly smaller than a typical error of distance modulus at redshifts $<$0.2. We used a diameter of $logd25$ because it is already in logarithmic scale. Because supergalactic coordinates 
are periodic, as are any spherical coordinates, we converted them to 3D Cartesians with unit radial vector for all galaxies.

Here, we considered 12 attributes of galaxies, which are the input explanatory variables to predict our desired target: distance modulus $m-M$.
We chose these variables because they are easily accessible for observation and are available for a large number of galaxies. 
In particular, galaxy coordinates provide information about the LSS, colour indices correlate with a morphological type, and
photometric data and angular diameters correlate with distance modulus. Therefore, we used a compilation of different parameters, some of which, such as for example position in the sky, surface brightness, and colour indices, have never been used before for distance modulus estimation. 

\section{Linear regression}
We explain the main principles of machine-learning regression on a linear model.
This is a basic regression model that deals with linear combinations of input variables (also referred to as features or attributes).
Multidimensional linear regression is a system that takes an n-size vector of input explanatory variables $\textit{\textbf{x}}\in R^n$
and predicts a scalar, so-called dependent variable, $y\in R,$ with some approximation $\hat{y}$:
\begin{equation}
    \hat{y}=\textit{\textbf{w}}^\top\textit{\textbf{x}}+b=w_{1}x_{1}+...+w_{n}x_{n}+b,
        \label{eq:lin}
\end{equation}
where the vector of model parameters is $\textit{\textbf{w}}\in R^{n}$ and $b \in R$ is the intercept term (bias).
Parameters $\textit{\textbf{w}}$ could be interpreted as the weights of the contribution of certain features to the composed output value $\hat{y}$. 
The larger the  absolute weight of a feature, $w_{i}$, the larger the impact of the $i$-th feature on the prediction.

The mean squared error (MSE) of the predicted output values $\hat{y}$ with respect to the real $y,$

\begin{equation}
    MSE=1/m \sum_{i}^{m}(\hat{y}_{i}-y_{i})^{2},
        \label{eq:mse}
\end{equation}
is widely used as an indicator of model performance; here $m$ is the size of the sample. 
In other words, the model performance could be expressed as the Euclidean distance between predicted $m$-dimension vectors $\hat{y}$ and $y$.

The main requirement of the machine learning regression is that the algorithm work well on new inputs that were not involved in the training stage.
Therefore, we split our data into training and test samples. Typically, the test sample represents a randomly selected 20-35\% subsample of the observations from the primary sample.
It is important that the training and test samples be independent and have the same unbiased distributions. 

The model should optimise its parameters on a training sample with minimisation $MSE_{train}$ and evaluate the obtained ($\textit{\textbf{w}}$, $b$) 
parameters on a test sample.
The test error $MSE_{test}$ should be minimised as well. 
This error with its variation reflects the expected level of error of $y$ for the new inputs $\textit{\textbf{x}}$.
Such a procedure is called `generalisation' and differentiates the machine learning technique from a simple fitting.
There are two main problems, which may appear during the training: underfitting, when the training error is too large, and overfitting, when the training error is small but the test error is still large \citep{Goodfellow2016}.

Linear regression is one of the simplest models, with a small capacity but a high parsimony and learning speed. 
For training of linear regression and all the models considered here, we used the free machine-learning library Scikit-learn\footnote{https://scikit-learn.org}.
To prevent very large parameters $\textit{\textbf{w}}$, we applied the Ridge regularisation \citep{Ng2004}, where the loss function 
for minimisation is $MSE_{train}+||\textit{\textbf{w}}||^{2}$.
We normalised all features before training by removing the mean and scaling to unit variance, which is a common requirement for many machine-learning estimators.
To evaluate the accuracy of our predictive model with the new data, we used the k-folds cross-validation technique \citep{Kohavi95astudy}:
we randomly split our sample into five equally sized subsamples and trained the model five times, sequentially using one subsample as the test sample and the other four as the training samples, rotating the test subsample each time. 
This gave us five independent estimations of $MSE_{test}$ from which we calculated the mean and its standard deviation.

For the larger $S_{0.50}$ sample, a linear regression gives the test root-mean-square error 
$RMSE_{test}=0.376 \pm 0.003$ mag. Using Eq.~\ref{eq:mM} this error was converted to a linear relative distance error of $17.3\%\pm0.2\%$.
We obtained the following coefficients of linear regression $\textit{\textbf{w}}$ in decreasing order of absolute value:
$w_{V_{LG}}=0.713$, $w_{K}=0.134$, $w_{logd25}= -0.110$, $w_{bri25}=0.103$, $w_{U}=-8.92$x$10^{-2}$, $w_{B-K}= 6.75$x$10^{-2}$, $w_{U-I}=6.37$x$10^{-2}$, $w_{Y_{SG}}=3.08$x$10^{-2}$, $w_{X_{SG}}= 1.25$x$10^{-3}$, $w_{Z_{SG}}=-7.45$x$10^{-5}$, $w_{B}=0$, $w_{I}=0$.
The most important parameter here is the radial velocity $V_{LG}$, and those with the least significance are the $K$ magnitude, angular diameter, and surface brightness.
The remaining parameters provide a smaller contribution to the distance estimation.

In cases where we do not have a redshift or $V_{LG}$ value, it is still possible to recover distance using other data 
with an accuracy of 0.52 mag or $24\%$  of the angular size distance. The errors for different models considered in this paper can be found in Table 1.
For the sample of nearby galaxies with more accurate measurements of distance modulus $S_{0.25}$, we obtained $RMSE_{test}=0.288\pm 0.019 $ mag 
($ 13\%\pm 0.9 \%$). In the case of radial velocity estimation, we obtained an error of $0.644\pm 0.029 $ mag ($ 30\%\pm 1.3 \%$).

\section{Polynomial regression}

A polynomial regression is an extension of a linear regression, where 
a  $k^{th}$ degree polynomial relation between the input explanatory variables $\textit{\textbf{x}}$ and the dependent variable $y$ is used.
We considered second- and third-order polynomial regressions.

For the second-order polynomial regression we obtained $RMSE_{test}=0.366 \pm 0.003$ mag 
($16.9\%\pm0.1\%$) for $S_{0.50}$ sample.
For the third-order, $RMSE_{test}^{3pol}=0.362 \pm 0.037$ mag ($16.7\%\pm1.7\%$).
As the third-order regression offers only a small improvement with respect to the second-order regression but with a greater uncertainty, we did not consider it.
For the $S_{0.25}$ sample we obtained an error of $0.276\pm 0.017 $ mag ($ 13\%\pm 0.9 \%$). 
When we eliminate the radial velocity, the test error is $0.607\pm 0.021 $ mag ($ 28\%\pm 1 \%$).

\section{A k-nearest neighbours regression}

\begin{figure}
        \includegraphics[width=\columnwidth]{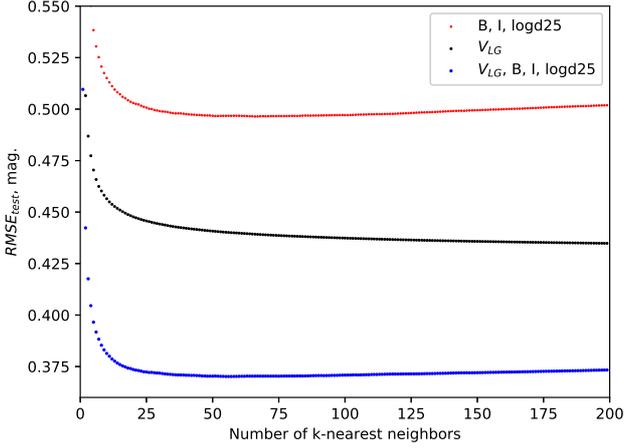}

    \caption{Dependence of $RMSE_{test}$ of k-nn regression on $k$ number of neighbours for different sets of input features: I, B bands and angular diameter (red dots), radial velocity (black), I, B bands, angular diameter, and radial velocity (blue).}
    \label{fig:knn}
\end{figure}

The k-nearest neighbours (k-NN) regression uses the $k$ closest training points around the test point in the feature $n$D space.
The predicted variable $\hat{y}$ in this case is the weighted average of the values $y$ of the k nearest neighbours \citep{Altman1992}. A
k-NN algorithm computes distances or similarities between the new instance and the training instances to make a decision.
Normally, it uses weighted averaging, where each neighbour among the closest $k$ neighbours has a weight of $1/d$, where $d$ is the distance to the neighbour.

k-NN regression is a type of instance-based learning.
Contrary to this approach, a linear regression and many other function-based approaches use explicit generalisation. 

k-NN regression has one hyperparameter, $k$, which allows the number of near neighbours to be taken into account.
We obtained our best results  for the $S_{0.50}$ sample for the case where distance weight to neighbour was inverse to the distance with Euclidean metric, with $k=56$ near neighbours.
We obtained an error of $RMSE_{test}=0.370 \pm 0.003$ mag ($17.0\%\pm0.1\%$) using radial velocity, angular diameters, and photometry in B and I bands. As can be seen from Fig.~\ref{fig:knn}, the minimum error forms a plateau in the range $40<k<80$.  
A combination of photometric data and angular diameter, simultaneously excluding radial velocity, provides an error of 0.50 ($23\%$).
The test errors of regression for the $S_{0.25}$ sample are listed in Table 1.

\section{Gradient-boosting regression}

The gradient-boosting regression is a kind of ensemble algorithm, which is widely used in machine learning. 
The ensemble algorithm is a stack of simple prediction models, like decision tree, linear regression, and so on,  joined together to make a final prediction. 
The theory behind this method is that many weak models predict a target variable with independent individual error.
Superposition of these models could show better results than any single predictor alone.

There are two main approaches of forming the ensemble algorithm; `bagging', where all simple models are independent and the final result is averaged over each model output.
The second approach is referred to as boosting, where the predictors are lined up sequentially and each subsequent predictor learns from the errors of the previous one, reducing these errors. We applied gradient-boosting regression \citep{mason99} using the open-source software library XGBoost\footnote{https://xgboost.ai}.
This algorithm minimises the error function by iteratively choosing a function that points to the negative gradient direction in the space of model parameters.

To prevent overfitting we applied
DART (Dropouts meet Multiple Additive Regression Trees) technique, which decreases the effect of overspecialisation when adding new trees. We found the optimal hyperparameters when using gradient-boosting regression and obtained a test error of $RMSE_{test}=0.355 \pm 0.003$ mag ($16.3\%\pm0.1\%$).

\section{Neural-network regression}

The Multilayer Perceptron is a type of feedforward ANN  consisting of neurons grouped in parallel layers: an input layer, a hidden layer or layers, and an output layer.
All the neurons of adjacent layers are connected. The main characteristic of an ANN is the ability to transmit a numerical signal from one artificial neuron to another 
in a `feedforward' direction from the input layer to the output layer.
In a regression model, the output layer is a single neuron that computes the target value $\hat{y}$.
The output of each neuron is transformed by some non-linear activation function of the sum of its inputs.
The outputs of the neurons from one layer are the input for neurons of the next layer, and so on. The connection between neurons has a weight that adjusts as learning proceeds. 
This weight corresponds to the importance of a signal at its transmission.

The neural network is a very powerful tool for machine-learning regression since it can approximate the 
continuous function of many variables with any accuracy under certain conditions \citep{Cybenko1989}. 
The ANN regression utilises a supervised learning technique called backpropagation of error (loss function) for its training.
In this study, we used the mean squared error (Eq. \ref{eq:mse}) as a loss function.

For our regression task we took all 12 input features mentioned in Sect. 2 as neurons of the input layer.
The best model performance was reached by shallow ANN with two hidden layers with 24 and 228 neurons, respectively.
As an activation function we used rectified linear unit function `relu', $f(x) = max(0, x)$.
The Broyden-Fletcher-Goldfarb-Shanno algorithm is considered to be the most appropriate for our task; it
is one of the quasi-Newton methods for solving ANN optimisation problems \citep{Curtis2015}.
The regularisation term of L2 penalty was chosen as 0.0005 to avoid over-fitting.
The learning rate was in mode ‘invscaling’.

\begin{figure}
        \includegraphics[scale=0.8]{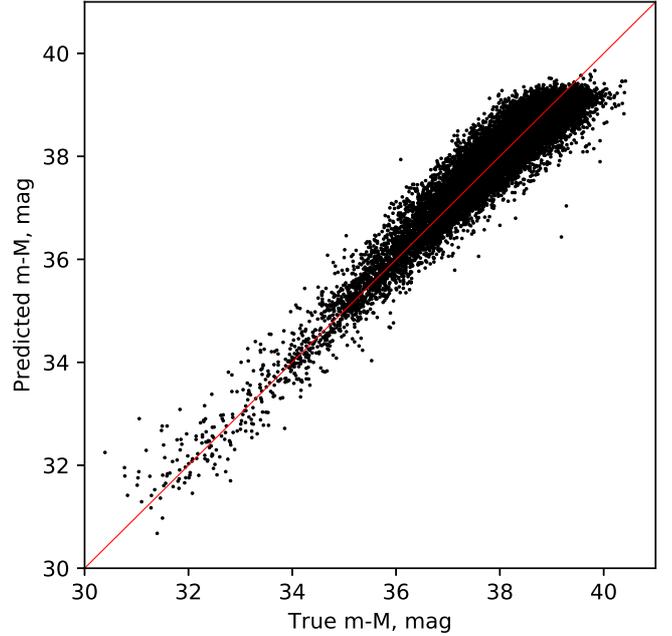}

    \caption{Dependence of the predicted distance moduli versus the true distance moduli for an ANN regression model using all the attributes for the $S_{0.5}$ sample.}
    \label{fig:real_predict}
\end{figure}

Finally, we obtained a test error $RMSE_{test}=0.354 \pm 0.003$ mag ($16.3\%\pm0.1\%$), which is comparable to the gradient-boosting result.
The plot of predicted versus true values is shown in  Fig.~\ref{fig:real_predict}.

\section{Discussion and conclusions}

\begin{table*}[t]
        \centering
        \caption{List of applied regression models with $RMSE_{test}$ errors using all the attributes of galaxies and without the $V_{LG}$ value. The last ten columns show the importance of each attribute for a given regression model, which is given as an increase in error as a percentage when leaving out a given attribute.}
        \label{tab:rezults}
        \begin{tabular}{l|cc|cccccccccc}

                Model      & all            &  without      & SGB      & $V_{LG}$  & $logd25$& $bri25$ & U       & B & I      & K      & U-I    & B-K   \\
                                   & attributes     &    $V_{LG}$   & SGL      &           &         &         &         &   &        &        &        &        \\

                \hline
                Linear     &  0.38          &  0.52          &  0.48    & 37       &  0.22   &    0.64 & 0.01    & 0 &   0    & 0.01   &  0.14  & 0.02     \\
                Polynomial    & 0.37           &   0.49         & 0.87     &   32     &   0.27  &  1.16   & 0.78    & 0.06 & 0      &  0.05  & 1.67   & 0.05  \\
                k-nn       &  0.37          &  0.50          &    0     &   35     &0.01     & 0       &  0      &0.01  &0.02    &  0     &  0     & 0     \\
                Gradient boosting  & 0.36           & 0.44           &    1.51      & 22         &  0.18       &    0     & 0.04        &1.02      &0.95        &    0       &0           &0      \\
                ANN       &0.35  & 0.44  &    1.09  &26&0.33&0&0.16&0.01&0.47&0.11&0.71&0.22\\
                \hline

                \end{tabular}
\end{table*}

Table 1 presents root-mean-square test errors for the sample $S_{0.5}$ (described in Section 2) after applying regression models for two sets of input attributes. The first set includes all attributes: photometric data, angular diameter, surface brightness, colour indices, radial velocity, and position of a galaxy in the sky (second column). The second set has the same attributes but without radial velocity (third column). The lowest errors are for the gradient-boosting and the neural network regressions. However, the ANN model has a smaller complexity than the gradient-boosting model and we chose ANN regression as the most appropriate model according to Occam's razor principle, i.e. ANN regression has a good trade-off between simplicity and accuracy.

The last ten columns of Table 1 show the importance of each observable galaxy attribute.
We quantified this importance as the increase in test error, as a percentage, due to leaving out a given attribute. The most important contribution comes from radial velocity, especially for linear, polynomial, and k-nn regressions (32-37\%). At the same time, a radial velocity is less important for the gradient-boosting and ANN models (22-26\%), as these approaches effectively also use information from other attributes. For ANN regression, the most important after $V_{LG}$ is Supergalactic coordinates of galaxy (1.09\%), $U-I$ colour (0.71\%), $I$ mag (0.47\%), and angular diameter (0.33\%). The rest of attributes individually have importance less than 0.22\%.


Table 2 shows the mean errors of various methods for $m-M$ computing applied to the $S_{0.50}$ sample using all attributes. 
As can be seen, the ANN regression has an error of 0.35 mag, which lies between the values of the errors of the 
BCG\footnote{A secondary distance indicator by the brightest galaxies in galaxy clusters as standard candles \citep{1980ApJ...241..493H}}
and TF relation\footnote{standard candles based on the absolute blue magnitudes of spiral galaxies \citep{1977A&A....54..661T}} 
methods, and shows a better result than the FP relation (0.42 mag).
Direct conversion of a radial velocity to $m-M$ according to Eqs. 1-3 gives a mean error of 0.40 mag (Conv. $V_{LG}\rightarrow$$m-M$ in Table 2).
Therefore, usage of all available attributes improves the uncertainty from 0.40 to 0.35 mag.

When we exclude $V_{LG}$ from consideration, our method gives an error of 0.44 mag (20\%), which is still acceptable for studying the LSS web and galaxy evolution. 
The case of using all the attributes but without radial velocity of galaxies is analogous to photometric redshift computation. The only difference is that the dependent variable is distance modulus instead of radial velocity. The relative RMSE test error of 20\% will also apply to the redshift predictions.
As can be seen from the second column of Table 2, the ANN regression (all attributes) could  potentially be applied to 393 359 galaxies, which is the number of unique galaxies from the Lyon-Meudon Extragalactic Database for which all ten observational parameters needed for this model are available. The catalogue of distance modulus \citep{2017AJ....153...37S} contains 183058 unique galaxies. Therefore, we expect to contribute an additional 210301 distance moduli computed with an accuracy of 0.35 mag at $z<0.2$.
Furthermore, the Lyon-Meudon Extragalactic Database contains 436 140 galaxies with nine of the required observational parameters but with unknown radial velocity.
We expect to compute $m-M$ with an accuracy of 0.44 mag, and redshift with an accuracy of 20\% for around 40 000 galaxies with unknown radial velocities.

\begin{table}
        \centering
        \caption{ Comparison of one-sigma statistical error of the distance modulus for different methods from the NASA/IPAC Extragalactic Database with three cases considered in our work: 
        the ANN regression using all attributes and without radial velocity, and using direct conversation radial velocity to $m-M$ according to Eq. 1-3.
        Errors are only for the $S_{0.50}$ sample ranked in ascending order separately for primary and secondary methods. The column N represents the number of galaxies for which distance modulus was defined by a given method from \cite{2017AJ....153...37S}. For the ANN regression and conversion of $V_{LG}$ to $m-M$ models we give the number of galaxies for which it is possible to compute distance modulus using the corresponding method. }
        \label{tab:err}
        \begin{tabular}{l|r|c} 

Method& N &error, mag\\
                \hline
                \multicolumn{3}{c}{Primary methods}\\
TRGB&   475&0.05\\
Cepheids&87     &0.08\\
PNLF    &72     &0.12\\
GC radius&      107&0.13\\
HII region diameter&44  &0.13\\
SNIa&   3179&0.14\\
SNIa SDSS&1771  &0.16\\
SNII optical&184&0.17\\
SBF     &539&0.18\\
AGN time lag&20&0.18\\
GCLF    &213&0.18\\
Masers  &10&0.22\\
BCG     &239&0.35\\
\multicolumn{3}{c}{Secondary methods}\\
Sosies  &344&0.20\\
Tertiary&283&0.30\\
D-Sigma &566&0.33\\
ANN regr. (all attributes)& 393359 & 0.35\\

TF &12244&    0.38\\
Conv. $V_{LG}$ to $m-M$ & 1209871&0.40 \\

FP&129038         &0.42\\
ANN regr. (without $V_{LG}$)&436140  & 0.44\\
                \hline
                \end{tabular}
\end{table}

\begin{figure}
        \includegraphics[width=\columnwidth]{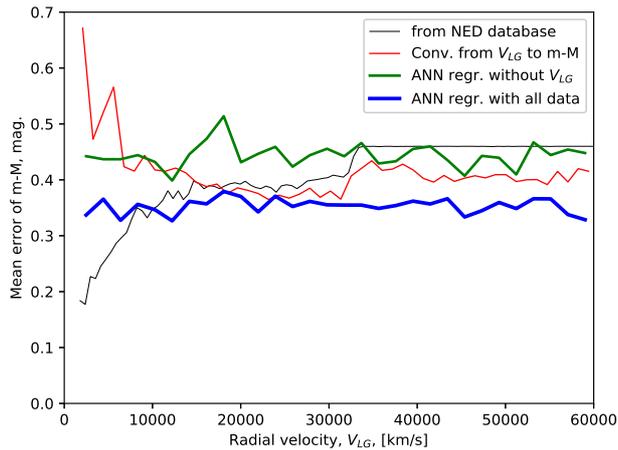}

    \caption{Root-mean-square errors of $m-M$ for various measurements of galaxies with different radial velocities. In order of line width increasing: 
    distance modulus statistical errors from the NASA/IPAC extragalactic database used in our study (black line),
    after conversion $V_{LG}$ to $m-M$ (red line), using the ANN regression without $V_{LG}$ (green line), using the ANN regression with all available 
    attributes (blue line).}
    \label{fig:err_vlg}
\end{figure}

We demonstrate how the ANN regression reconstructs $m-M$ at different distances in  Fig.~\ref{fig:err_vlg}.
The black line represents one-sigma statistical error from the NASA/IPAC extragalactic database for galaxies from the $S_{0.5}$ sample.
The error increases from 0.2 to 0.4 at radial velocities below 10000 km/s. 
This can be explained by the decreased contribution of high-accuracy Cepheid and RR Lyrae measurements of $m-M$.
The error reaches 0.46 mag for galaxies with $V_{LG}>30000~$km/s. 
The majority of the distance moduli for those galaxies were estimated using the FP method with an error of 0.46 mag. 

Direct use of the radial velocity as an analogue of linear distance according to Eqs. 1-3 causes large errors of 0.5-0.6 mag 
for $V_{LG}<7000~$km/s due to the influence of collective motions caused by local galaxy clusters and voids.
At larger distances, this kind of distance measurement provides almost constant error of around 0.40 mag.

Our ANN regression method, even without information about $V_{LG}$, could improve the accuracy of distance measurements for nearby galaxies and provides RMSE 0.44 mag for all distances.
Addition of the radial velocity to our ANN regression model decreases the error to 0.35 mag for all distances.
The performance of our ANN model demonstrates the fact that the influence
of velocity at small radial velocities should be decreased in favour of other parameters such as
angular diameter, photometry, and so on. 
This is why for our model the influence of local clusters and voids is negligible and the distance modulus error is almost constant at 0.35 mag for all $V_{LG}$ (blue thick line Fig.~\ref{fig:err_vlg}).

Our approach is especially useful for measuring distances for galaxies with $V_{LG}>10000~$km/s, where primary methods are not working.
Therefore, the regression model developed here is competitive against widely used secondary methods for $m-M$ measurement such as the FP and TF relation.

We proposed the new data-driven approach for computing distance moduli to local galaxies based on ANN regression.
In addition to traditionally used photometric data we also involved the surface brightness, angular size, radial velocity, and the position in the sky of galaxies to predict $m-M$.
Applying our method to the test sample of randomly selected galaxies from the NASA/IPAC Extragalactic Database we obtained a root-mean-square error of 0.35 mag (16\%),
which does not depend on the distance to a galaxy and is comparable with the mean errors from the TF and FP methods. Our model provides an error of 0.44 mag (20\%) in cases where radial velocity is not taken into consideration.

In the future we plan to make a more detailed comparison of our ANN regression approach with other physically based methods.
Namely, we are going to build an analogue of the 2D redshift space correlation function with distance modulus instead of redshift.
More accurate methods should reveal the smaller effects of distortion caused by both random peculiar velocities of galaxies and by the coherent motions of galaxies in the LSS.
Also, we will apply our ANN model to measure $m-M$ for galaxies with unknown distance moduli in the range of radial velocities $1500<V_{LG}<60000~$km/s 
and release a corresponding catalogue. This catalogue will include distance moduli computed for around a quarter of a million galaxies with $z<0.2$.

\begin{acknowledgements}
We would like to thank very much referee for kind advices and useful remarks. This work was partially supported in frame of the budgetary program of the NAS of Ukraine "Support for the development of priority fields of scientific research" (CPCEL 6541230).    
\end{acknowledgements}

\bibliographystyle{aa} 
\bibliography{references} 

\end{document}